\documentclass{article}
\usepackage{arxiv}
\usepackage[utf8]{inputenc} 
\usepackage[T1]{fontenc}    
\usepackage{hyperref}       
\usepackage{url}            
\usepackage{booktabs}       
\usepackage{amsfonts}       
\usepackage{amsmath}
\usepackage{nicefrac}       
\usepackage{microtype}      
\usepackage{cleveref}       
\usepackage{graphicx}
\usepackage{natbib}
\usepackage{doi}
\usepackage[title]{appendix}

\newcommand{\E}{\mathbb{E}}
\newcommand{\Var}{\mathrm{Var}}
\newcommand{\Cov}{\mathrm{Cov}}
\newcommand{\N}{\mathcal{N}}

\usepackage{booktabs}
\usepackage{rotating}
\usepackage{siunitx}

\title{Ensemble-size-dependence of deep-learning post-processing methods that minimize an (un)fair score: motivating examples and a proof-of-concept solution}

\author{ \href{https://orcid.org/0000-0002-2958-6637}{\includegraphics[scale=0.06]{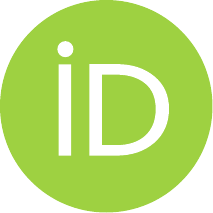}\hspace{1mm}Christopher David Roberts}\\
	ECMWF\\
	Shinfield Park\\
	Reading, United Kingdom\\
	\texttt{chris.roberts@ecmwf.int} \\
}
\renewcommand{\headeright}{PREPRINT}
\renewcommand{\undertitle}{PREPRINT}
\renewcommand{\shorttitle}{Ensemble-size-dependence of deep-learning post-processing methods}

\hypersetup{
  colorlinks   = true, 
  urlcolor     = blue, 
  linkcolor    = blue, 
  citecolor   = red 
}

\begin{document}
\maketitle

\begin{abstract}
Fair scores reward ensemble forecast members that behave like samples from the same distribution as the verifying observations. They are therefore an attractive choice as loss functions to train data-driven ensemble forecasts or post-processing methods when large training ensembles are either unavailable or computationally prohibitive. The adjusted continuous ranked probability score (aCRPS) is a fair and ensemble-size-unbiased scoring rule when applied to ensembles with forecast members that are interpretable as conditionally independent draws from an underlying predictive distribution. However, distribution-aware post-processing methods that introduce structural dependency between members can violate this assumption, such that aCRPS no longer functions as a fair scoring rule. We demonstrate this effect using two approaches designed to minimize the expected aCRPS of a finite ensemble: (1) a linear member-by-member calibration, which couples members through a common dependency on the sample ensemble mean, and (2) a deep-learning method, which couples members via transformer self-attention across the ensemble dimension. In both cases, the results are sensitive to ensemble size and apparent gains in aCRPS can correspond to systematic unreliability characterized by over-dispersion. We introduce trajectory transformers as a proof-of-concept that ensemble-size independence can be achieved. This approach is an adaptation of the Post-processing Ensembles with Transformers (PoET) framework and applies self-attention over lead time while preserving the conditional independence required by aCRPS. When applied to weekly mean $T_{2m}$ forecasts from the ECMWF subseasonal forecasting system, this approach successfully reduces systematic model biases whilst also improving or maintaining forecast reliability regardless of the ensemble size used in training (3 vs 9 members) or real-time forecasts (9 vs 100 members).

\end{abstract}

\section{Introduction}
\label{section:intro}

Ensemble forecasting can be viewed as a Monte Carlo method, in which ensemble members approximate samples from a probability distribution of future atmospheric states, such that each member represents an equally plausible forecast trajectory \citep{leith1974theoretical,molteni1996ecmwf,leutbecher2008ensemble}. In practice, forecast ensemble members are generated through a combination of initial-condition perturbations and stochastic parameterisations \citep{leutbecher2008ensemble,lock2019treatment}. For a given start date, the resulting members can to a good approximation be treated as independent realisations from an underlying predictive distribution that is itself conditional on the observed initial state\footnote{Forecast members aggregated across many start dates can be correlated through their common dependence on the initial conditions. Throughout this paper, statements that ensemble members are `independent draws from a predictive distribution' should be understood as conditional independence given the initial state.}. However, state-of-the-art forecast models may suffer from systematic biases and flow-dependent errors due to many factors, including discretization and parameterization errors, unresolved or simplified physical processes, and errors in the specified initial state \citep{bauer2015quiet, magnusson2019dependence}. Raw ensemble forecasts can thus be biased and unreliable, where reliability is a measure of the statistical consistency between forecast probabilities and observed frequencies \citep{wilks2011statistical, gneiting2007probabilistic}. 

In a statistical sense, the goal of ensemble forecasting is to maximize forecast sharpness subject to reliability \citep{gneiting2007probabilistic}. This principle has motivated the use of proper scoring rules and their `fair' variants \citep{gneiting2007probabilistic, ferro2014fair} as loss functions for ensemble post-processing methods \citep{gneiting2005calibrated, rasp2018neural, gronquist2021deep, bouallegue2024improving} and data-driven ensemble forecast systems \citep{lang2024aifs, kochkov2024neural}. A negatively oriented scoring rule is considered strictly proper if the expected score is uniquely minimized when the predictive distribution is equal to the true distribution of the observations \citep{gneiting2007strictly}. Fair scores are proper scoring rules that account for finite ensemble size effects and reward forecast members that behave as though they are sampled from the same distribution as the verifying observations \citep{ferro2014fair}. 

The distributional form of the continuous ranked probability score (CRPS) is a strictly proper score \citep{gneiting2007strictly} that is defined as the integral
\begin{equation}
    \label{eq:crps_integral}
    \mathrm{CRPS}_{\mathrm{int}}(F, y) \;=\; \int_{-\infty}^{\infty} \bigl(F(x) - \mathbf{1}\{x \ge y\}\bigr)^2 \, dx,
\end{equation}
where $F$ represents a forecast cumulative distribution function and $\mathbf{1}\{x \ge y\}$ is the indicator function, equal to $1$ when $x \ge y$ and $0$ otherwise. \citet{gneiting2007strictly} showed that the CRPS can be equivalently defined using a kernel representation

\begin{equation}
    \label{eq:crps_kernel}
    \mathrm{CRPS}_{\mathrm{ker}}(F, y) \;=\; \E_F|X - y| \,-\, \tfrac{1}{2}\,\E_F|X - X'|,
\end{equation}
where $X$ and $X'$ are independent copies of a random variable with predictive distribution $F$ and finite first moment. Under these conditions, equation \ref{eq:crps_kernel} is exactly equal to equation \ref{eq:crps_integral} and inherits its strict propriety. For a finite ensemble with members $\{x_1, \ldots, x_N\}$ and observation $y$, the unadjusted sample estimate of the CRPS is defined by 

\begin{equation}
    \label{eq:CRPS}
    \mathrm{CRPS} = \frac{1}{N}\sum_{k=1}^{N}|x_k - y| - \frac{1}{2N^2}\sum_{k=1}^{N}\sum_{l=1}^{N}|x_k - x_l|.
\end{equation}
Although this definition uses the kernel representation, it is mathematically identical to the equation derived by substituting the empirical CDF $\hat{F}(x) = \frac{1}{N}\sum_{k=1}^N\mathbf{1}\{x \ge x_k\}$ into equation \ref{eq:crps_integral} \citep{zamo2018estimation}. This equivalence of the empirical-CDF and kernel sample estimators means that they share the same independence conditions required to target the proper score $\mathrm{CRPS}_{\mathrm{int}}$. However, equation \ref{eq:CRPS} is biased for finite ensembles and rewards overconfident forecasts. To address this limitation, \citet{ferro2008effect} defined an adjusted version of the sample CRPS, which can be written in the kernel representation following \citet{leutbecher2019ensemble} as

\begin{equation}
    \label{eq:aCRPS}
    \mathrm{aCRPS} = \frac{1}{N}\sum_{k=1}^{N}|x_k - y| - \frac{1}{2N(N-1)}\sum_{k=1}^{N}\sum_{\substack{l=1 \\ l \neq k}}^{N}|x_k - x_l|.
\end{equation}
When forecast members can be interpreted as independent and identically distributed (iid) samples from an underlying predictive distribution, equation \ref{eq:aCRPS} is both \emph{fair} and unbiased with respect to ensemble size \citep{ferro2014fair}. When these conditions are not met, aCRPS can no longer be applied as a fair scoring rule at finite ensemble sizes. For example, consider a situation in which members are identically distributed but not necessarily independent. The second term on the right hand side of equation \ref{eq:aCRPS} then provides an unbiased estimate of $\tfrac{1}{2}\E_F|X_1 - X_2|$, where $X_1$ and $X_2$ are exchangeable in the sense that they share the same marginal distribution $F$ but may be correlated. If members are not independent, $\E_F|X_1 - X_2|$ is not representative of $\E_F|X - X'|$ and the resulting score does not target the proper score given by equation \ref{eq:crps_integral}.

The difference between these terms can be defined $\Delta \equiv \E_F|X_1 - X_2| - \E_F|X - X'|$, such that $\Delta=0$ when forecast members are iid. By substitution into equation \ref{eq:crps_kernel} it can be shown that the unadjusted and adjusted sample estimators given by equations \ref{eq:CRPS} and \ref{eq:aCRPS} include terms proportional to $\Delta$, with coefficients $\frac{N-1}{2N}$ and $\frac{1}{2}$ respectively. The larger weighting of $\Delta$ in the aCRPS sample estimator means that dependence between members at small ensemble sizes can produce large biases relative to the proper integral score, in some cases exceeding the corresponding bias of the unadjusted sample CRPS estimator. Concrete examples of these effects are illustrated in section \ref{section:idealized_example}.

Fair scoring rules are thus specific to the dependence structure of the sampled ensemble members and do not exist for all forms of dependency \citep{ferro2014fair}. For this reason, distribution-aware post-processing methods that introduce structural dependency\footnote{Note that structural dependency is distinct from more general statistical dependency and is used throughout this paper to refer to coupling between members that is induced by a post-processing architecture, such as a shared dependence on the sample ensemble mean.} between ensemble members can break the underlying assumptions of an otherwise fair score (e.g. aCRPS) such that members are rewarded when they appear to be sampled from a different distribution to the verifying observations. In this case, the post-processed forecasts that minimize the target loss function become systematically unreliable such that forecasts and observations have different statistical properties. This type of unreliability is commonly diagnosed from a mismatch between the average ensemble variance and the average squared error of the ensemble mean (after correction for finite ensemble size effects) or from differences in the total variance of forecast members and observations when evaluated over many events \citep{leutbecher2008ensemble, johnson2009reliability, roberts2025unbiased}. 

Traditional post-processing methods can be broadly categorized into two classes: (1) parametric methods, which assume a specific distributional form (e.g., Gaussian) and estimate distribution parameters from ensemble statistics \citep{gneiting2005calibrated, scheuerer2015probabilistic}, and (2) non-parametric methods, which adjust each ensemble member individually and can preserve multivariate ensemble dependencies \citep{van2015ensemble, scheuerer2015variogram}. Recent advances in machine learning methods, combined with the increasing accessibility of deep learning software frameworks and large reforecast datasets, have also facilitated the development of sophisticated data-driven ensemble post-processing methods, which are able to learn complex non-linear relationships from high-dimensional input data \citep{gneiting2005calibrated, rasp2018neural, gronquist2021deep, bouallegue2024improving, horat2024deep}. However, the computational expense of generating datasets and training deep-learning methods means that data-driven post-processing methods are commonly trained using a small ensemble size (e.g. 10 members) with the goal of applying these methods to real-time forecasts with larger ensemble size (e.g. 50+ members). For this reason, it is important to demonstrate that the benefits of a data-driven post-processing method trained at one ensemble size are preserved when it is applied at a larger ensemble size.  

\begin{figure}[!htbp]
    \includegraphics[width=14cm]{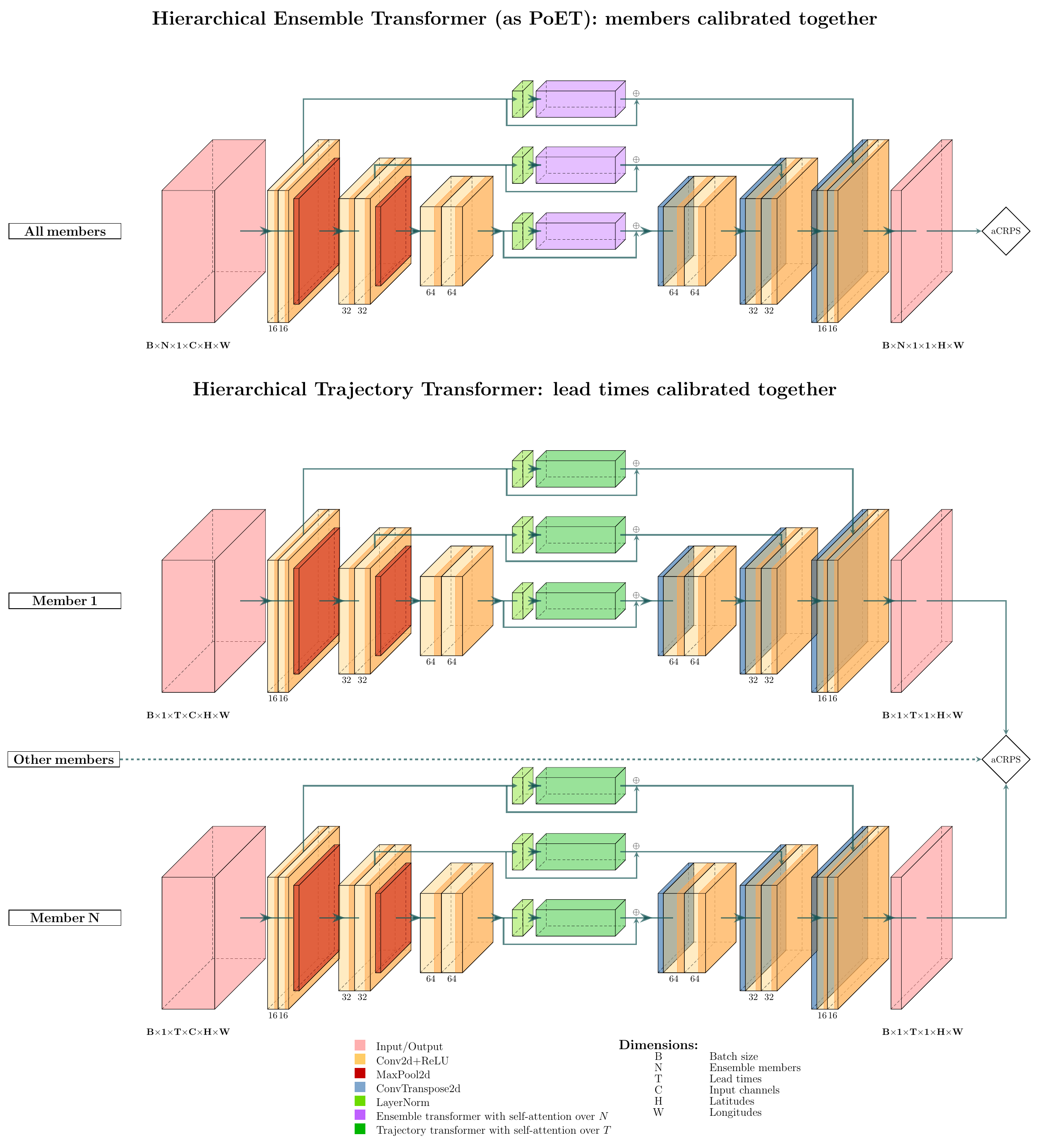}
    \centering
    \caption{(Top) Schematic diagram of the PoET hierarchical ensemble transformer architecture as configured during training \citep{bouallegue2024improving}. (Bottom) Schematic diagram of the hierarchical trajectory transformer described in this study, as configured during training. Figure generated using PlotNeuralNet \citep{iqbal_plotneuralnet_2018}.    
    }
    \label{fig:schematic}
\end{figure}

This study focuses on a new class of ensemble post-processing methods that leverage transformer architectures \citep{vaswani2017attention}, which have been demonstrated to outperform other deep-learning solutions when applied to medium-range weather forecasts \citep{bouallegue2024improving}. Specifically, we consider the Post-processing Ensembles with Transformers (PoET) framework, which combines a hierarchical encoder-decoder convolutional U-net architecture with \emph{ensemble transformer} processor blocks \citep[figure \ref{fig:schematic}; ][]{finn2021self}. The application of the transformer self-attention mechanism across the ensemble dimension allows for information exchange between ensemble members during post-processing, and is intended to allow the model to learn location- and member-specific corrections after drawing context from the entire forecast distribution. 

The ensemble transformer has several properties that are desirable for a post-processing method. Firstly, the distribution-aware nature of the method means that the sample ensemble mean and variance are adjusted directly based on learned relationships from past forecasts. Secondly, the ensemble transformer can be considered a type of member-by-member method as it outputs post-processed ensemble members, which can be evaluated and integrated into downstream applications the same way as raw forecasts. Lastly, the transformer architecture is ensemble-size agnostic, in the sense that the attention matrix is generated dynamically and thus the number of ensemble members used in training and inference may differ. However, this architecture also has two important disadvantages. Firstly, the ensemble transformer method is not trajectory-aware, such that lead times are calibrated separately without considering temporal relationships within the same forecast. Secondly, the exchange of information between members via self-attention injects structural dependency, which can result in systematic unreliability when combined with sample CRPS loss functions (equations \ref{eq:CRPS} and \ref{eq:aCRPS}).

To demonstrate that ensemble-size independence is achievable while preserving the conditional independence required for aCRPS to be a fair training objective, we introduce the \emph{trajectory transformer} as a proof-of-concept. This revised approach retains the hierarchical encoder-decoder convolutional U-net architecture from PoET but applies the transformer self-attention mechanism across the forecast lead-time dimension independently to each member (figure \ref{fig:schematic}). This approach is thus trajectory-aware and can be considered a form of lead-time-continuous post-processing, which has recently been investigated using an Ensemble Model Output Statistics (EMOS) method \citep{wessel2024lead}. This change in architecture provides two important benefits: (i) it provides the opportunity to learn physically meaningful spatio-temporal relationships, including lagged error structures, and (ii) zero information exchange between members during inference, which provides an ensemble-size-independent approach to forecast post-processing that is suitable for use with aCRPS as a loss function. The main disadvantage of this method is that ensemble members must be updated without direct knowledge of the full sample distribution, which means corrections to the ensemble mean and variance must be learned implicitly. In this paper we illustrate the ensemble-size-independence of the trajectory transformer approach when applied to weekly mean data from the European Centre for Medium-Range Weather Forecasts (ECMWF) subseasonal prediction system.

The remainder of this paper is organized as follows: Section \ref{section:idealized_example} uses idealized Gaussian data and a linear member-by-member calibration to illustrate the concept of architecture-induced dependence between members as a source of miscalibration when optimizing sample CRPS estimators. Section \ref{section:methods} describes the ECMWF subseasonal forecast training and evaluation data and the transformer-based post-processing methods used in this study. Section \ref{section:results} evaluates the performance of the ensemble transformer and trajectory transformer methods applied to ECMWF subseasonal forecasts of weekly mean two-metre temperature (T$_{2m}$). Finally, section \ref{section:conclusions} provides a discussion of the results and summarizes the main conclusions of this study.

\section{An idealized example: Gaussian forecasts and linear calibration}
\label{section:idealized_example}
This section uses idealized Gaussian forecast data to illustrate the impact on reliability of a linear member-by-member calibration that minimizes the expected value of either CRPS or aCRPS. Let $\{x_{k,j}\}$ represent an idealized ensemble forecast of \(k=1,\dots,N\) members covering \(j=1,\dots,M\) independent cases with $\{y_j\}$ verifying observations, generated by a signal-plus-noise model

\begin{align}
    \label{eq:gaussian_data}
    x_{k,j} &= s_j + n_{k,j}, \\
    y_j &= s_j + e_j,
\end{align}
where $s_j$, $n_{k,j}$, and $e_j$ are mutually independent with $s_j \sim \N(0, \sigma_s^2)$, $n_{k,j} \sim \N(0, \alpha^2)$, and $e_j \sim \N(0, \beta^2)$. These forecasts are unbiased but potentially unreliable. In this study, we diagnose reliability using two metrics that measure statistical consistency between forecasts and observations in expectation across independent cases $j = 1, \ldots, M$. Firstly, we define a sample estimator for the total variance ratio (VR) as 

\begin{equation}
      \label{eq:VR}
      \mathrm{VR} = \frac{\frac{1}{N}\sum^{N}_{k=1} \sum^{M}_{j=1} \left( x_{k,j} - \left< x_k \right> \right) ^2}{\sum^{M}_{j=1} \left( y_j - \left< y \right> \right) ^2},
\end{equation}
where $\left<\cdot\right> \equiv \frac{1}{M}\sum^{M}_{j=1}(\cdot)_j$ denotes the mean over the $M$ cases. The variance ratio $\mathrm{VR} \to 1$ as $M \to \infty$ provided that the average marginal variance of forecast members across cases is equal to the variance of observations. It is thus possible for an ensemble forecast with dependency between members to still have $\mathrm{VR} \to 1$. We describe forecasts with $\mathrm{VR} > 1$ as over-active and those with $\mathrm{VR} < 1$ as under-active, according to whether the total variance of the forecast members is larger or smaller than that of the observations. Secondly, we define a sample estimator for the spread-error ratio (SER)  as 

\begin{equation}
      \label{eq:SER}
       \textnormal{SER} =  \sqrt{ \frac{N+1}{N-1}} \sqrt{ \frac{\sum^{M}_{j=1} \frac{1}{N}\sum^{N}_{k=1}\left(x_{k,j} - \bar{x}_j\right)^2}{\sum^{M}_{j=1}\left( y_{j} - \bar{x}_j \right)^2 } },
\end{equation} 
where the factor of $\frac{N+1}{N-1}$ ensures estimates are unbiased with ensemble size \citep{leutbecher2008ensemble} and $\bar{x}_j \equiv \frac{1}{N}\sum^{N}_{k=1}x_{k,j}$ denotes the sample ensemble mean over the $N$ members of case $j$. Provided that the average expected ensemble variance and average expected squared error of the ensemble mean are consistent with one another, $\mathrm{SER} \to 1$ as $M \to \infty$. We describe forecasts with $\mathrm{SER} > 1$ as over-dispersive and those with $\mathrm{SER} < 1$ as under-dispersive, according to whether the ensemble spread is larger or smaller than the typical error of the ensemble mean. For the idealised stationary Gaussian forecasts described above, $\mathrm{SER} \to 1$ requires that forecast members $x_{k,j}$ and observations $y_j$ are interpretable as independent draws from a common conditional distribution given $s_j$. If forecast members are exchangeable but correlated after conditioning on $s_j$ (e.g. due to a shared error common to all members), then $\mathrm{SER} \to 1$ is only possible when the forecast marginal distribution differs from the observed marginal distribution such that $\mathrm{VR} \neq 1$. For the more general situation where forecast distributions vary from case-to-case, $\mathrm{SER} \to 1$ is a necessary but not sufficient condition for reliability as it is possible for there to be compensation between different regimes that leaves the average value of SER unchanged.

Now consider a simple member-by-member linear calibration \citep{van2015ensemble}:
\begin{equation}
    \label{eq:mbm_calibration}
    \hat{x}_{k,j} = a + b\bar{x}_j + c(x_{k,j} - \bar{x}_j), 
\end{equation}

where $\hat{x}_{k,j}$ represents the calibrated ensemble member. In the case of an unbiased and perfectly reliable forecast (i.e. $\alpha = \beta$), we expect estimates of $a \to 0$ and $b,c \to 1$ as $M \to \infty$. This is achieved when parameters are estimated by enforcing reliability constraints (i.e. $\mathrm{VR}=1$, $\mathrm{SER}=1$) that are unbiased with ensemble size \citep{roberts2025unbiased}. However, this is not the case for finite $N$ when the objective is to minimize either $\E[\mathrm{CRPS}]$ or $\E[\mathrm{aCRPS}]$. For finite ensemble sizes, calibrated members share a common dependency on $\bar{n}_{j}$, allowing the adjusted CRPS of an already reliable small ensemble to be further reduced through adjustments to the variance of the ensemble mean and member perturbations. These apparent gains in forecast skill arise from exploiting the shared dependence between ensemble members and thus reflect a reduction in reliability rather than a genuine improvement in forecast quality. For the idealized Gaussian forecasts described above, the optimal values for $a$ and $b$ are the same for both adjusted and unadjusted CRPS objective functions and given by

\begin{equation}
    \label{eq:optimal_b_maintext}
    a^* = 0, \quad b^* = \frac{N\sigma_s^2}{N\sigma_s^2 + \alpha^2}.
\end{equation}

The optimal values for $c$ are given by
\begin{equation}
    \label{eq:optimal_c_maintext}
    c^*_{\mathrm{CRPS}} = \sqrt{\frac{N}{N+1}} \cdot \frac{\hat{\sigma}_\epsilon^*}{\alpha}, \quad c^*_{\mathrm{aCRPS}} = \frac{N}{\sqrt{(N-1)(N-2)}} \cdot \frac{\hat{\sigma}_\epsilon^*}{\alpha},
\end{equation}
where
\begin{equation}
    \hat{\sigma}_\epsilon^{*2} = \frac{\alpha^2\sigma_s^2}{N\sigma_s^2 + \alpha^2} + \beta^2.
\end{equation}

From these expressions it is clear that, for finite $N$, minimizing $\E[\mathrm{aCRPS}]$ produces calibrated forecasts with systematically higher variance than minimizing $\E[\mathrm{CRPS}]$. Furthermore, the optimal value for $c$ is undefined for $N < 3$ when minimizing $\E[\mathrm{aCRPS}]$. Further details on the derivation of these expressions are provided in Appendix A. The optimal linear calibration parameters, expected scores, and reliability properties of post-processed Gaussian forecasts are summarized in table \ref{tab:calib_all} for a range of ensemble sizes and the two CRPS-based objective functions.

\begin{figure}[!htbp]
    \includegraphics[width=15cm]{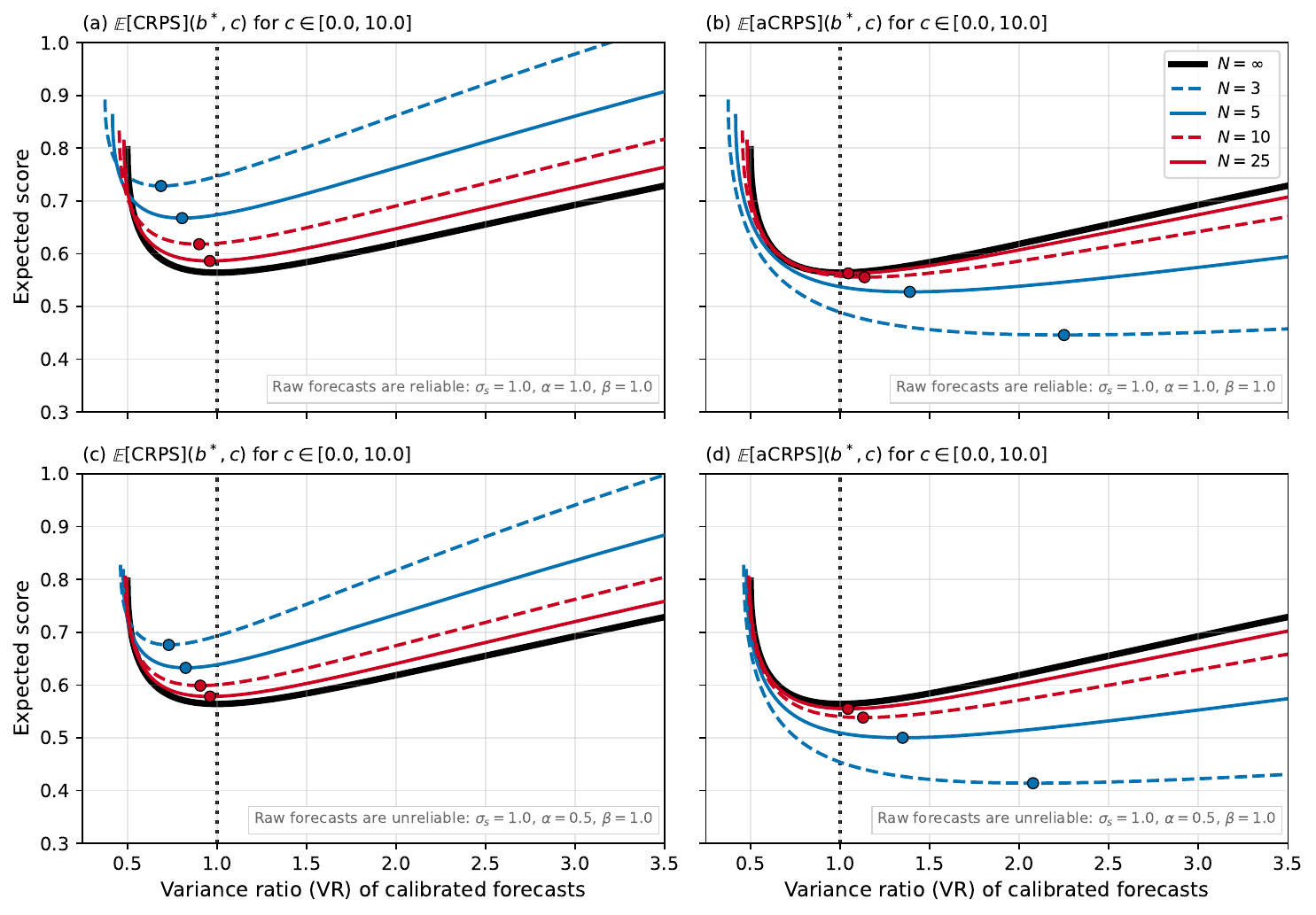}
    \centering
    \caption{Expected values of the unadjusted (left column) and adjusted (right column) CRPS sample estimators plotted against the total variance ratio (VR) of linearly-calibrated idealised Gaussian forecasts of different ensemble sizes as described in section \ref{section:idealized_example}. The expected scores are calculated using the optimal value of $b^*$ (equation \ref{eq:optimal_b_maintext}) and for a range of values of $c$, with optimal values of $c^*$ indicated as markers corresponding to the minima of the expected scores. The underlying raw forecasts are either reliable (top row) or unreliable (bottom row) and use model parameters consistent with table \ref{tab:calib_all}. Vertical dotted lines indicate the expected variance ratio of a reliable forecast.}
    \label{fig:expected_score_vs_var_ratio}
\end{figure}

When the underlying forecasts are statistically reliable ($\alpha=\beta$), both objectives yield optimal calibration parameters that converge $b^* \to 1$ and $c^* \to 1$ as $N \to \infty$. However, for finite ensembles the reliability properties of the resulting post-processed forecasts are very different. Minimization of $\E[\mathrm{CRPS}]$ preserves the desired spread-error properties of a reliable forecast (i.e. $\mathrm{SER}=1$), but reduces the total variance of the calibrated members such that forecasts are over-active (table \ref{tab:calib_all} and figure \ref{fig:expected_score_vs_var_ratio}a). In contrast, minimization of $\E[\mathrm{aCRPS}]$ encourages a large increase in forecast spread and total variance such that resulting post-processed forecasts are both over-active and over-dispersive (table \ref{tab:calib_all} and figure \ref{fig:expected_score_vs_var_ratio}b).

A similar effect is evident when the underlying raw forecasts are unreliable due to under-dispersion ($\alpha < \beta$), with $\E[\mathrm{CRPS}]$ minimization producing under-active forecasts (table \ref{tab:calib_all} and figure \ref{fig:expected_score_vs_var_ratio}) and $\E[\mathrm{aCRPS}]$ minimization producing over-active and over-dispersive forecasts (table \ref{tab:calib_all} and figure \ref{fig:expected_score_vs_var_ratio}d). Together, these results illustrate that neither CRPS nor aCRPS can be fair when applied to post-processed forecasts that introduce structural dependency between members. In this example, the aCRPS objective introduces a systematic bias towards over-active and over-dispersive ensembles, whereas the unadjusted CRPS rewards under-active forecasts. Furthermore, as highlighted in section \ref{section:intro}, these sample estimators respond asymmetrically to dependence between members such that minimizing $\E[\mathrm{aCRPS}]$ may result in a more substantial degradation of reliability (in terms of VR and SER) than minimizing $\E[\mathrm{CRPS}]$ when applied to small ensembles (table \ref{tab:calib_all} and figure \ref{fig:expected_score_vs_var_ratio}). 

These results are specific to the linear calibration scenario described above, but can be considered representative of the class of post-processing methods for which dependency between members vanishes as $N \to \infty$ (i.e. $\Delta \to 0$ in the notation of section \ref{section:intro}). This is the reason why the minimized values of $\E[\mathrm{aCRPS}]$ at finite ensemble size are not representative of the (higher) values that are achieved as $N \to \infty$ (figure \ref{fig:expected_score_vs_var_ratio}b and table \ref{tab:calib_all}). Post-processing methods that do not share these dependency characteristics may exhibit qualitatively different results.

\begin{figure}[!htbp]
    \includegraphics[width=8cm]{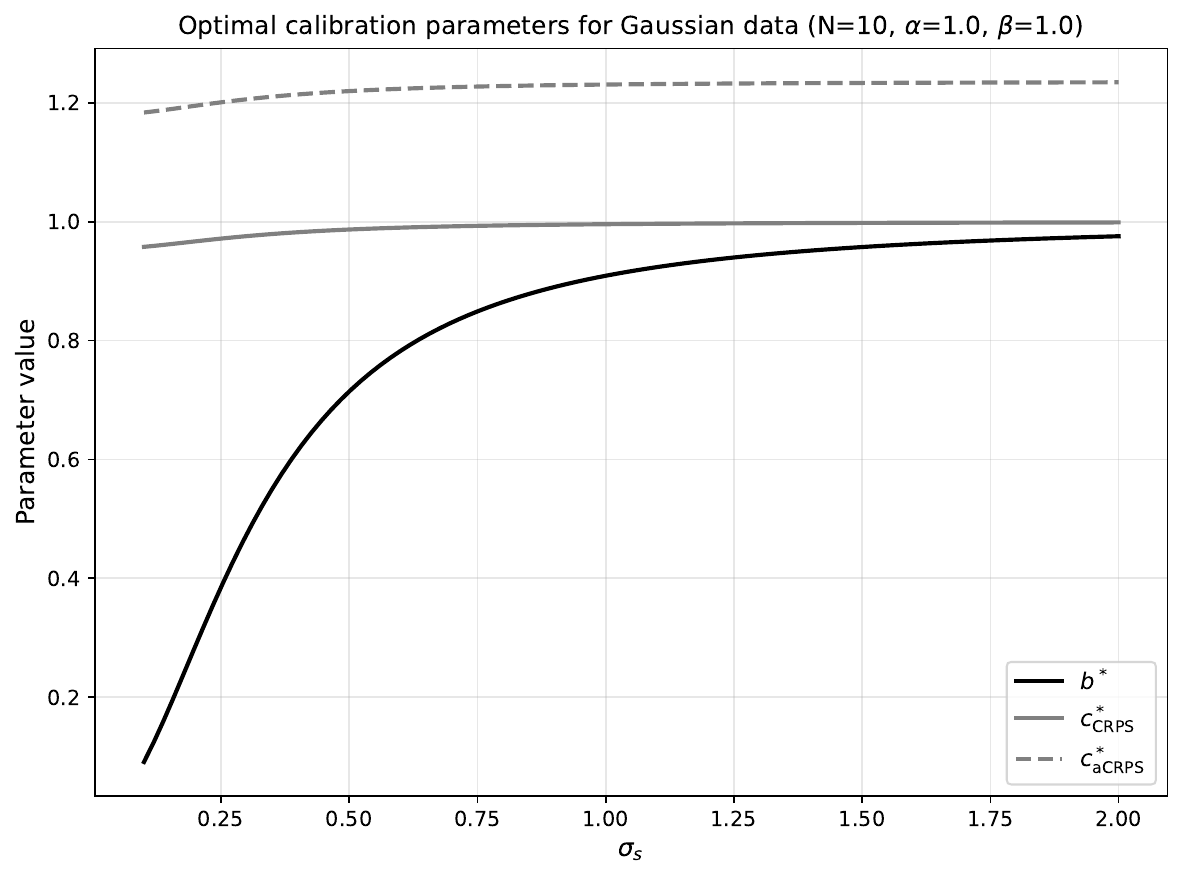}
    \centering
    \caption{Optimal linear calibration parameters for idealized Gaussian forecasts following equations \ref{eq:optimal_b_maintext} and \ref{eq:optimal_c_maintext} for a 10-member perfectly reliable ($\alpha = \beta$) forecast and different values of $\sigma_s$.}
    \label{fig:optimal_params}
\end{figure}

Figure \ref{fig:optimal_params} illustrates the sensitivity of these effects to changes in predictability in a perfectly reliable scenario, which can be approximated by variations in $\sigma_s$ for fixed $\alpha = \beta$. For modest ensemble sizes ($N=10$) and higher predictability (i.e. $\sigma_s > \alpha, \beta$), minimizing $\E[\mathrm{CRPS}]$ produces slightly under-active forecasts whilst minimizing $\E[\mathrm{aCRPS}]$ produces over-active and over-dispersive forecasts. For distribution-aware post-processing methods applied to high predictability regimes, minimization of $\E[\mathrm{CRPS}]$ rather than $\E[\mathrm{aCRPS}]$ may be a justifiable compromise if reliability-enforcing alternatives are not available. However, for lower predictability scenarios (i.e. $\sigma_s < \alpha, \beta$), there is no good compromise and the choice is between very under-active forecasts by minimizing $\E[\mathrm{CRPS}]$ or strongly over-dispersive and over-active forecasts by minimizing $\E[\mathrm{aCRPS}]$. These effects are therefore a practical concern for post-processing of (lower predictability) extended-range forecasts and also evident for the ensemble transformer post-processing method that introduces structural dependency between members through the self-attention mechanism across the ensemble dimension (section \ref{section:results}).

\begin{sidewaystable}[p]
    \centering
\caption{Optimal linear calibration parameters, reliability metrics (SER, VR), and expected scores for idealized Gaussian forecasts as a function of ensemble size $N$ following equations \ref{eq:optimal_b_maintext} and \ref{eq:optimal_c_maintext}. Note that SER is calculated from population parameters to be consistent with equation \ref{eq:SER} and therefore represents the spread-error ratio that would be be achieved with an infinite ensemble using the calibration parameters estimated from $N$ members.}
\label{tab:calib_all}
\small
\setlength{\tabcolsep}{4pt}
\renewcommand{\arraystretch}{1.15}
\begin{tabular}{
c
S S S S S S
S S S S S S
}
\toprule
& \multicolumn{6}{c}{Calibration minimizes $\mathbb{E}[\mathrm{CRPS}]$}
& \multicolumn{6}{c}{Calibration minimizes $\mathbb{E}[\mathrm{aCRPS}]$} \\
\cmidrule(lr){2-7} \cmidrule(lr){8-13}
{$N$}
& {$b^*$} & {$c^*$} & {SER} & {VR} & {$\mathbb{E}[\mathrm{CRPS}]$} & {$\mathbb{E}[\mathrm{aCRPS}]$}
& {$b^*$} & {$c^*$} & {SER} & {VR} & {$\mathbb{E}[\mathrm{CRPS}]$} & {$\mathbb{E}[\mathrm{aCRPS}]$} \\
\midrule
\multicolumn{13}{c}{Raw forecast data are reliable ($\sigma_s=\alpha=\beta=1$).} \\
\midrule
3    & 0.7500 & 0.9682 & 1.0000 & 0.6875 & 0.7284 & 0.5463 & 0.7500 & 2.3717 & 2.4495 & 2.2500 & 0.8921 & 0.4460 \\
10   & 0.9091 & 0.9959 & 1.0000 & 0.9008 & 0.6180 & 0.5619 & 0.9091 & 1.2309 & 1.2360 & 1.1364 & 0.6250 & 0.5556 \\
50   & 0.9804 & 0.9998 & 1.0000 & 0.9800 & 0.5754 & 0.5641 & 0.9804 & 1.0410 & 1.0412 & 1.0212 & 0.5756 & 0.5639 \\
100  & 0.9901 & 1.0000 & 1.0000 & 0.9900 & 0.5698 & 0.5642 & 0.9901 & 1.0203 & 1.0203 & 1.0103 & 0.5699 & 0.5641 \\
1000 & 0.9990 & 1.0000 & 1.0000 & 0.9990 & 0.5648 & 0.5642 & 0.9990 & 1.0020 & 1.0020 & 1.0010 & 0.5648 & 0.5642 \\
\midrule
\multicolumn{13}{c}{Raw forecast data are unreliable ($\sigma_s=1$, $\alpha=0.5$, $\beta=1$).} \\
\midrule
3    & 0.9231 & 1.7974 & 1.0000 & 0.7308 & 0.6761 & 0.5070 & 0.9231 & 4.4028 & 2.4495 & 2.0769 & 0.8280 & 0.4140 \\
10   & 0.9756 & 1.9300 & 1.0000 & 0.9069 & 0.5989 & 0.5445 & 0.9756 & 2.3856 & 1.2360 & 1.1280 & 0.6057 & 0.5384 \\
50   & 0.9950 & 1.9852 & 1.0000 & 0.9803 & 0.5712 & 0.5600 & 0.9950 & 2.0671 & 1.0412 & 1.0209 & 0.5715 & 0.5598 \\
100  & 0.9975 & 1.9926 & 1.0000 & 0.9901 & 0.5677 & 0.5621 & 0.9975 & 2.0330 & 1.0203 & 1.0102 & 0.5678 & 0.5620 \\
1000 & 0.9998 & 1.9993 & 1.0000 & 0.9990 & 0.5645 & 0.5640 & 0.9998 & 2.0033 & 1.0020 & 1.0010 & 0.5645 & 0.5640 \\
\bottomrule
\end{tabular}
\end{sidewaystable}

\section{Methods}
\label{section:methods}

\subsection{Transformer-based post-processing methods}
This study explores the impact of two transformer-based post-processing methods: (1) the hierarchical \emph{ensemble transformer} as described by \citet{bouallegue2024improving} and (2) a modification to this approach, that we term the hierarchical \emph{trajectory transformer}. Both methods utilize the PoET software framework\footnote{\url{https://github.com/ecmwf-lab/poet/}} and consist of an encoder-processor-decoder U-net structure.  The encoder and decoder layers consist of a combination of 2D convolutions and ReLU activations combined with max pooling and transpose convolutions, respectively, with processors consisting of a combination of layer normalization and transformer blocks (figure \ref{fig:schematic}).

The core of the original PoET architecture is the ensemble transformer, which was originally described by \citet{finn2021self}. The key idea is the application of the transformer self-attention mechanism \citep{vaswani2017attention} across the ensemble dimension, which provides an opportunity for each member to be modified based on context from an arbitrary number of other members. Details of the PoET implementation of the ensemble transformer are provided in Appendix B. The trajectory transformer approach used in this study uses the same attention and convolutional operations as the ensemble transformer, but applied along a different tensor dimension. In this implementation, the hidden state entering the trajectory transformer block at a given level is represented by
\begin{equation}
\label{eq:hidden_state_trajectory}
\mathbf{X} \in \mathbb{R}^{B \times T \times C \times H \times W},
\end{equation}
where $B$ is the batch size, $C$ is the channel depth, $H,W$ are spatial dimensions, and $T$ is the number of forecast lead times and replaces the ensemble dimension $N$ in equations \ref{eq:value_projection}--\ref{eq:output_projection}, while all channel and spatial operations remain unchanged. In this case, different forecast lead times may attend to one another and the transformed values for step $t$ (replacing member $i$) represent the original value combined with the attention-weighted combination of other lead times relative to the trajectory-mean value (i.e. mean over lead times for the given member). This structure provides an opportunity to learn physically meaningful lagged relationships while processing members entirely independently. 

Unlike ensemble members, forecast lead time is an ordered coordinate and therefore not permutation invariant. For this reason lead time is provided as a spatially uniform input feature, which acts as a form of positional encoding. This design ensures that ensemble members do not exchange information during inference, which preserves the conditional independence of members required for aCRPS to be fair. In principle this framework could be extended to other lead times, though this is not explored here.

\subsection{Transformer model training}

The ensemble and trajectory transformer methods are trained using identical network configurations (table \ref{tab:hyperparameters}) and the same 18 input fields (table \ref{tab:input_parameters}) to predict weekly mean T$_{2m}$ derived from ERA5 \citep{hersbach2020era5} on a 2.5$\times$2.5 regular latitude-longitude grid. Input fields were selected based on their ready availability as derived weekly means from a previous verification exercise. Both models use the Adam optimizer with an initial learning rate of $10^{-3}$ and a reduce-on-plateau scheduler combined with the adjusted CRPS loss function (equation \ref{eq:aCRPS}).  

During training, the ensemble transformer operates on all members together and each lead time is processed separately. In contrast, the trajectory transformer is applied separately to each member and aCRPS loss is aggregated over lead times (with equal weight) before back-propagation (figure \ref{fig:schematic}). The main difference during training is the use of a smaller batch size (16 vs 64) for the trajectory transformer, which is a consequence of the increased memory usage of the current implementation that loads all members and lead times together. Both models are trained for 100 epochs using the training period 1959-2017, with separate validation (2021-2023) and test (2018-2020) periods. To evaluate the sensitivity to ensemble size, we train each transformer method using nine members and again with a subset of three members that are randomly selected each batch. 

\begin{table}[!htbp]
    \centering
    \caption{Training and architecture hyperparameters for ensemble transformer and trajectory transformer models based on the PoET framework \citep{bouallegue2024improving}.}
    \label{tab:hyperparameters}
    \small
    \begin{tabular}{lcc}
    \toprule
    \textbf{Parameter} & \textbf{Ensemble Transformer} & \textbf{Trajectory Transformer} \\
    \midrule
    \multicolumn{3}{l}{\textit{Architecture}} \\
    Attention dimension & Ensemble members ($N$) & Lead times ($T$) \\
    Encoder channels & [16, 32, 64] & [16, 32, 64] \\
    Transformers per level & 1 & 1 \\
    Attention heads & 8 & 8 \\
    Input channels & 22 & 22 \\
    Output channels & 1 & 1 \\
    \midrule
    \multicolumn{3}{l}{\textit{Training}} \\
    Loss function & aCRPS &  aCRPS \\
    Batch size & 64 & 16 \\
    Initial learning rate & $10^{-3}$ & $10^{-3}$ \\
    Optimizer & Adam & Adam \\
    LR scheduler & ReduceLROnPlateau & ReduceLROnPlateau \\
    LR reduction factor & 0.3 & 0.3 \\
    LR patience (epochs) & 10 & 10 \\
    Minimum learning rate & $5\times10^{-5}$ & $5\times10^{-5}$ \\
    Training epochs & 100 & 100 \\
    \midrule
    \multicolumn{3}{l}{\textit{Data}} \\
    Training period & 1959--2017 & 1959--2017 \\
    Validation period & 2021--2023 & 2021--2023 \\
    Test period & 2018--2020 & 2018--2020 \\
    Ensemble size & 9 & 9 \\
    Lead times (hours) & [168, 336, 504, 672, 840, 1008] & [168, 336, 504, 672, 840, 1008] \\
    Grid resolution & 2.5$^\circ$ $\times$ 2.5$^\circ$ & 2.5$^\circ$ $\times$ 2.5$^\circ$ \\
    \bottomrule
    \end{tabular}
\end{table}

\begin{table}[!htbp]
    \centering
    \caption{Input features for post-processing weekly mean T$_{2m}$ forecasts. Forecast fields are standardized to zero mean and unit variance with precipitation log-epsilon transformed prior to scaling.}
    \label{tab:input_parameters}
    \small
    \begin{tabular}{lp{10cm}}
    \toprule
    \textbf{Category} & \textbf{Description} \\
    \midrule
    \textit{Forecast fields (18 channels)} & 2-metre temperature \\
    & Temperature at 850, 500, 200, and 50 hPa \\
    & Geopotential height at 500 hPa \\
    & Mean sea-level pressure \\
    & Total precipitation rate \\
    & 10-metre zonal wind \\
    & 10-metre meridional wind \\
    & Zonal wind at 850, 500, 200, and 50 hPa \\
    & Meridional wind at 850, 500, 200, and 50 hPa \\
    \textit{Static fields (2 channels)}  & Land-sea mask (0--1) \\
    & Topography (0--1) \\
    \textit{Auxiliary fields (2 channels)} &  Forecast lead time (0--1) \\
    & Top-of-atmosphere incoming solar radiation \\
    \bottomrule
    \end{tabular}
\end{table}

\subsection{Subseasonal ensemble forecasts and verification data}
The transformer methods described above are applied to weekly-mean data derived from 46-day forecasts produced with cycle 47R3 of the European Centre for Medium-Range Weather Forecasts (ECMWF) Integrated Forecasting System (IFS). We use data from a reduced-resolution configuration ($\sim$50 km  atmosphere, $\sim$100 km ocean) consisting of a 9-member extended reforecast dataset initialized on the 1st, 8th, 15th, and 22nd of each month over the period 1959-2023, yielding a total of 3120 forecast initializations. This reforecast dataset is described in detail by \citet{roberts2023euro} and \citet{roberts2025ensemble}. To assess the sensitivity of post-processing performance to ensemble size, we additionally use 100-member forecasts from the same IFS configuration, run for selected start dates (1$^{st}$ February, May, August, and November) during the period 2018-2023. Ensemble spread in coupled forecasts is generated through a combination of initial condition perturbations and stochastic parameterizations in the atmospheric model as described in \citet{lock2019treatment}. Forecasts are verified against weekly mean T$_{2m}$ data from the ERA5 analysis \citep{hersbach2020era5}.

\section{Impact of transformer-based post-processing methods}
\label{section:results}
\begin{figure}[!htbp]
    \includegraphics[width=12cm]{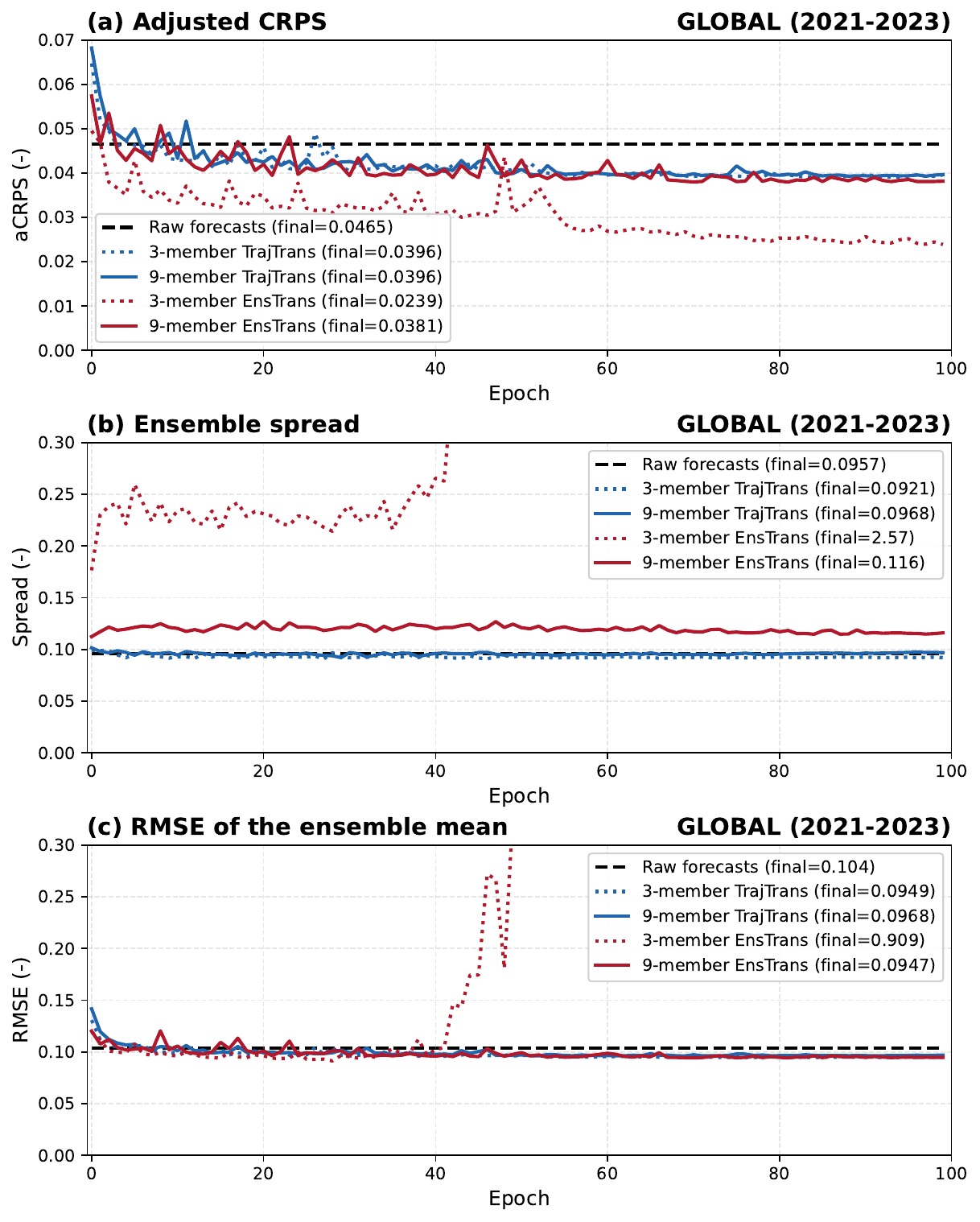}
    \centering
    \caption{(a) Weighted global mean aCRPS of raw and post-processed T$_{2m}$ forecasts in normalized units as a function of training epoch for the independent evaluation period (2021-2023). Each epoch represents a single pass through the entire training dataset (1959-2017). (b-c) As above, but for unbiased estimates of the ensemble spread and root mean square error (RMSE) of the ensemble mean calculated to be unbiased with ensemble size following \citet{leutbecher2008ensemble} and \citet{roberts2025unbiased}. All metrics are averaged over lead times with equal weight.}
    \label{fig:training_metrics}
\end{figure}

Figure \ref{fig:training_metrics} shows the evolution of evaluation metrics from the validation data (2021-2023) as a function of training epoch for ensemble and trajectory transformers, both of which are trained and evaluated for two different ensemble sizes. In all cases, aCRPS is reduced below the reference value of the uncalibrated raw forecasts. For the trajectory transformers, there is very close agreement between models trained with either three or nine ensemble members. In both cases, aCRPS and RMSE are reduced and the close agreement between (unbiased) estimates of ensemble spread and RMSE is consistent with a reliable forecast (i.e. $\mathrm{SER}=1.0$).

In contrast, the ensemble transformer method shows large differences between models trained using different ensemble sizes. When trained with nine members, the ensemble transformer appears to have a slightly lower aCRPS than both trajectory transformers. However, this apparent improvement in skill comes at the cost of reliability such that spread is $\sim$20 \% higher than RMSE ($\mathrm{SER}=1.2$). Furthermore, when trained with three members, the ensemble transformer method produces much lower aCRPS but the spread and RMSE are unstable and reach values that are both extremely over-dispersive ($\mathrm{SER}=2.8$) and an order of magnitude larger than the raw forecasts. These results are consistent with the theoretical results presented in section \ref{section:idealized_example} and illustrate that aCRPS cannot be fair, and is therefore not a meaningful metric of forecast quality, in the presence of conditional dependency between members introduced by post-processing methods. These results are robust across validation and test periods (not shown). The remainder of this paper focuses on transformer models trained on 9-members and evaluated using the combined validation-test period (2018-2023) to increase the sample size for more robust regional statistics.

\begin{figure}[!htbp]
    \includegraphics[width=11cm]{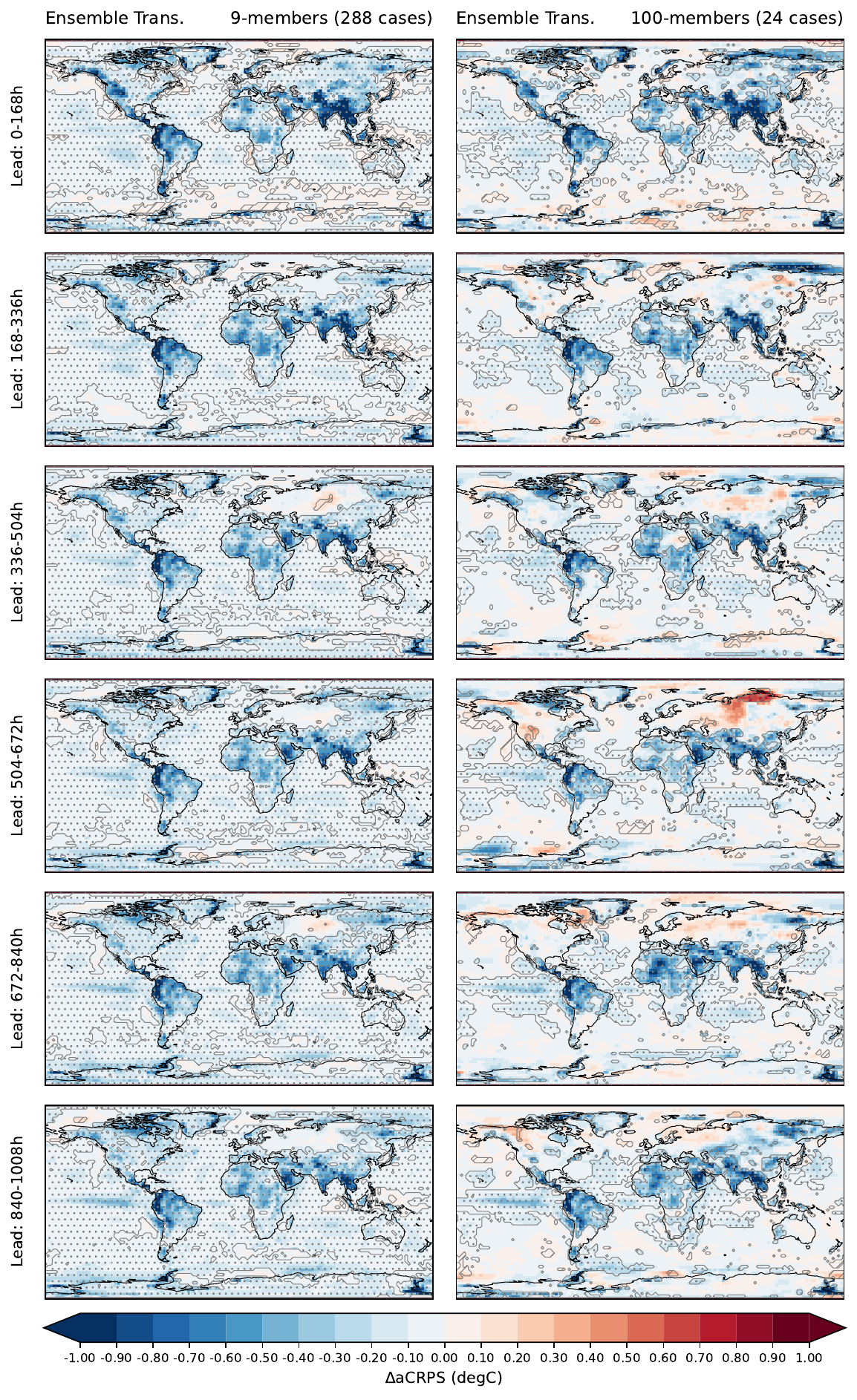}
    \centering
    \caption{Impact of PoET hierarchical Ensemble Transformer post-processing method \citep{bouallegue2024improving} on the adjusted CRPS of weekly mean T$_{2m}$ forecasts as a function of forecast lead time, where negative values are indicative of lower aCRPS in the post-processed reforecasts. The ensemble transformer post-processing method was trained on 9-members (see figure \ref{fig:training_metrics}) for the period 1959-2017 and evaluated on either 9-member (left) or 100-member (right) forecasts for the period 2018-2023. Stippling indicates regions where 95 \% confidence intervals for $\Delta$aCRPS do not include zero. Confidence intervals are estimated as the 2.5 and 97.5 percentiles of an empirical distribution derived by block bootstrap resampling (with replacement) 500 times. Start dates within the same calendar month are treated as a single block during resampling to preserve temporal correlation structure. }
    \label{fig:enstrans_fcrps_maps}
\end{figure}

\begin{figure}[!htbp]
    \includegraphics[width=12cm]{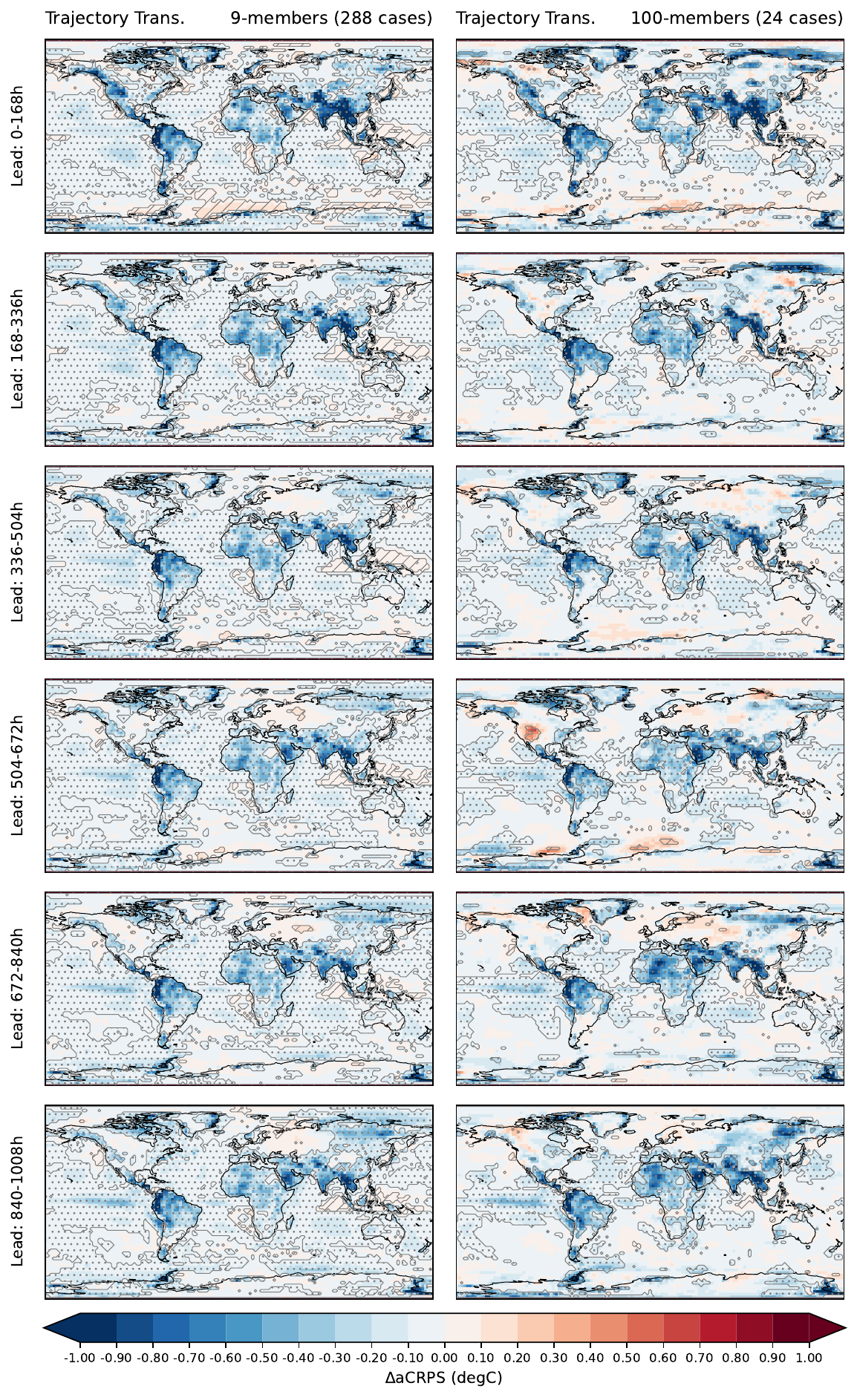}
    \centering
    \caption{As figure \ref{fig:enstrans_fcrps_maps}, but for the hierarchical Trajectory Transformer approach described in this study.}
    \label{fig:trajtrans_fcrps_maps}
\end{figure}

The regional impacts of transformer-based T$_{2m}$ post-processing methods relative to the uncalibrated raw forecasts are illustrated for different lead times in figures \ref{fig:enstrans_fcrps_maps} and \ref{fig:trajtrans_fcrps_maps}. For both post-processing methods, changes in aCRPS are quantified for two different datasets: (i) the validation-test set of 288  start dates of 9-member forecasts, which match the ensemble size used during model training, and (ii) a subset of 24 start dates of 100-member forecasts produced with the same model but different initial condition perturbations. To first order, the regional changes in aCRPS are similar across all lead times and all four combinations of ensemble size and post-processing method. These changes mostly represent real improvements to the mean state that are well-captured by both transformer-based methods. These effects are most pronounced over land, where the underlying IFS forecasts suffer from systematic biases in T$_{2m}$ relative to the ERA5 target. 

\begin{figure}[!htbp]
    \includegraphics[width=14cm]{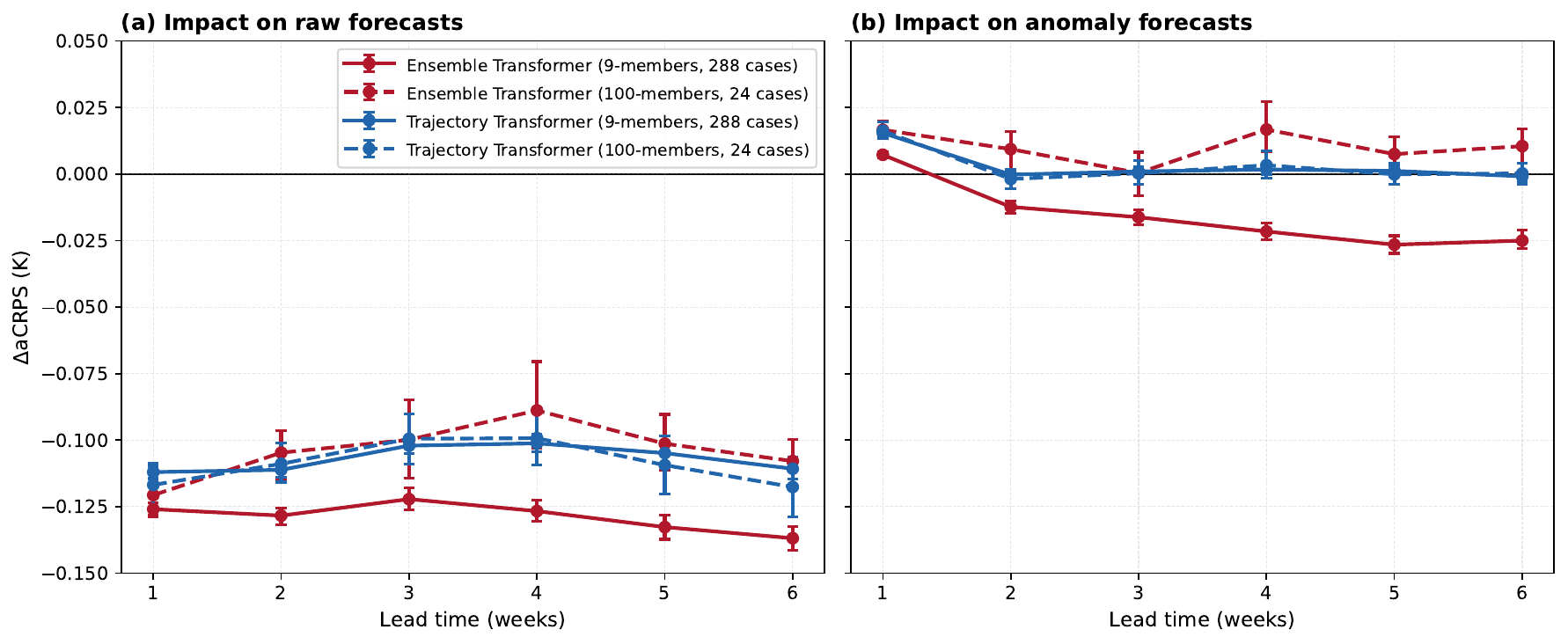}
    \centering
    \caption{(a) Impact of transformer-based post-processing methods on weighted global mean aCRPS calculated from weekly mean T$_{2m}$, where negative values are indicative of lower aCRPS in the post-processed reforecasts. (b) As (a), but for weekly mean forecast anomalies calculated following the `by-member--other-years' method of \citet{roberts2025unbiased}. Error bars correspond to 95 \% confidence intervals estimated as the 2.5 and 97.5 percentiles of an empirical distribution derived by block bootstrap resampling (with replacement) 500 times. Start dates within the same calendar month are treated as a single block during resampling to preserve temporal correlation structure.}
    \label{fig:global_fcrps}
\end{figure}

The regional impacts on aCRPS are superficially consistent between different ensemble sizes, though the reduced sample size of the 100-member forecasts reduces the number of comparisons that are robust to sampling uncertainties. However, on closer inspection, systematic differences between the ensemble and trajectory transformer approaches, including their sensitivity to ensemble size, become more apparent. The trajectory transformer approach is independent of ensemble size such that the impact on global aCRPS is extremely similar when evaluated using either 9-member or 100-member forecasts (figures \ref{fig:trajtrans_fcrps_maps} and \ref{fig:global_fcrps}a). This ensemble-size-independence is evident when evaluating trajectory-transformed outputs either as raw values or anomalies calculated with respect to start-date and lead-time dependent climatologies (figure \ref{fig:global_fcrps}b).

\begin{figure}[!htbp]
    \includegraphics[width=14cm]{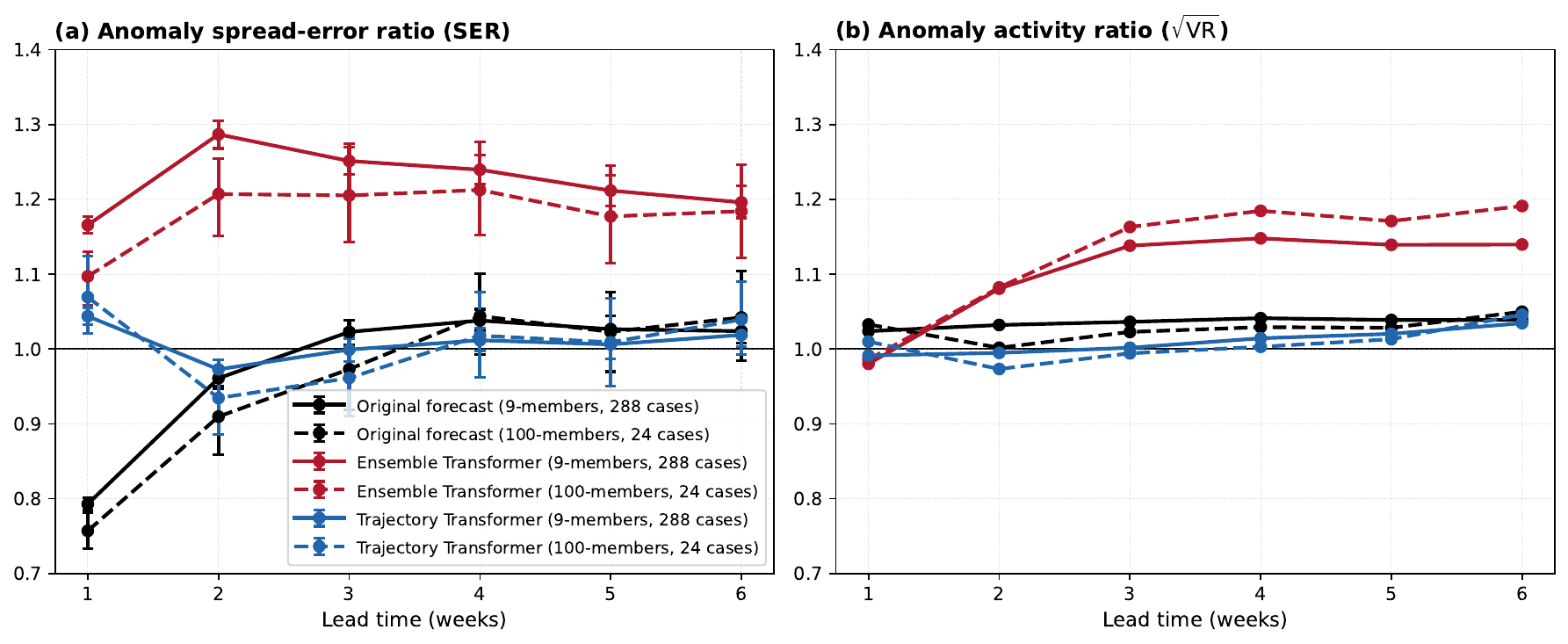}
    \centering
    \caption{(a) Anomaly-based global spread-error ratios (SER) calculated following equation 34 of \citet{roberts2025unbiased} combined with their `by-member--other-years' method for climatology calculation such that estimates are unbiased with respect to both ensemble size and climatology sample size. Error bars correspond to 95 \% confidence intervals estimated as the 2.5 and 97.5 percentiles of an empirical distribution derived by block bootstrap resampling (with replacement) 500 times. Start dates within the same calendar month are treated as a single block during resampling to preserve temporal correlation structure. Anomaly-based global activity ratios defined as the square root of a generalisation of equation \ref{eq:VR} such that the numerator and denominator represent the weighted global mean variance of weekly mean anomalies in forecast members and observations, respectively. Confidence intervals were not estimated for this quantity.}
    \label{fig:sea_ratio}
\end{figure}

As highlighted above, the positive impacts of the trajectory transformer post-processing method are dominated by the improvements to the mean state and the benefits are limited after accounting for the changes in the mean state. In fact, when evaluated as anomalies, week 1 post-processed forecasts are slightly degraded relative to the uncalibrated forecasts. This likely represents some degree of overfitting to the training data associated with non-stationarity of the underlying forecasts and associated initial condition uncertainties. However, there are some improvements to the reliability of the trajectory-transformed anomalies as diagnosed using spread-error ratios and total variance ratios (figure \ref{fig:sea_ratio}). Crucially, these results are independent of both the ensemble size used to train the trajectory transformer (figure \ref{fig:training_metrics}) and the ensemble size of the real-time forecast to be calibrated (figures \ref{fig:trajtrans_fcrps_maps}-\ref{fig:sea_ratio}).

In contrast, although the ensemble transformer successfully reduces mean T$_{2m}$ biases, the results are not independent of ensemble size and training with 9-members produces forecasts that are systematically unreliable with $\mathrm{SER} > 1$ and $\mathrm{VR} > 1$ (figure \ref{fig:sea_ratio}). Furthermore, the architecture-induced dependency between ensemble members means that the evaluation of 9-members provides an overoptimistic impression of the performance with 100-members (figure \ref{fig:global_fcrps}). In this example, the 9-member aCRPS results could be interpreted naively to  indicate that the ensemble transformer produces forecast anomalies that are significantly improved at weeks 2-6 compared to both the trajectory transformer and the original uncalibrated forecasts. However, this reduction of aCRPS does not generalise to the 100-member forecasts, which are seemingly degraded relative to the uncalibrated forecast anomalies. These results highlight the dangers of using sample CRPS estimators to evaluate post-processed forecasts that break the underlying assumption that members are conditionally independent draws from a predictive distribution.

\section{Discussion and conclusions}
\label{section:conclusions}

This study has highlighted a fundamental limitation of distribution-aware forecast post-processing methods that introduce structural dependency between members when minimizing the expected value of sample CRPS estimators. In the case of the adjusted CRPS (equation \ref{eq:aCRPS}), such methods break the underlying assumptions that make the score fair such that members are rewarded when they appear to be sampled from a different distribution than the verifying observations.  In this scenario, post-processed forecasts become systematically unreliable but appear more skillful when evaluated using aCRPS in both training and independent test datasets. We illustrate these effects using two approaches: (i) a simple linear member-by-member calibration, for which a theoretical explanation is provided (section \ref{section:idealized_example}), and (ii) the hierarchical ensemble transformer, as implemented in the PoET framework \citep{bouallegue2024improving}.


The key feature of the ensemble transformer method \citep{finn2021self} is the application of a self-attention mechanism across the ensemble dimension, such that each member is updated using context from an arbitrary number of other members. However, this exchange of information injects structural dependency between members, which results in ensemble size sensitivity during training and systematic unreliability when combined with sample CRPS loss functions such as equations \ref{eq:CRPS} and \ref{eq:aCRPS}. These effects are illustrated in section \ref{section:results}, where the PoET ensemble transformer is used to post-process weekly mean $T_{2m}$ forecasts from the ECMWF IFS subseasonal forecasting system with the objective of minimizing $\E[\mathrm{aCRPS}]$. Although the ensemble transformer post-processing method effectively reduces systematic model biases and achieves lower aCRPS values, the resulting post-processed forecasts are systematically unreliable, with spread approximately 20\% higher than RMSE when trained with 9 members. Furthermore, the 9-member evaluation of forecast anomalies provides an overoptimistic impression of the performance with 100 members, which end up degraded relative to uncalibrated forecast anomalies.

Although the structural dependency introduced by the ensemble transformer is more complex than the linear calibration of section \ref{section:idealized_example}, these methods share an analagous mechanism. Specifically, the transformed value of each member is updated by an attention-weighted sum of ensemble member value perturbations relative to the sample ensemble mean (equation \ref{eq:transformed_value} in appendix \ref{appendix:ensemble_transformer}). This dependence on an ensemble mean quantity is conceptually similar to the $\bar{x}_j$ dependence in the linear calibration, and likely explains why the ensemble transformer exhibits a similar ensemble size sensitivity to the idealized example.

This study introduces an alternative ensemble-size-independent approach to forecast post-processing that we term the trajectory transformer. This revised approach retains the same hierarchical encoder-decoder convolutional U-net architecture of PoET but applies the transformer self-attention mechanism across the forecast lead-time dimension independently to each member (figure \ref{fig:schematic}). This architectural change provides two key benefits: (i) it preserves the per-start-date independence of ensemble members (conditional on the initial state), ensuring compatibility with fair scoring rules and (ii) it provides an opportunity to learn physically meaningful spatio-temporal relationships and lagged error structures across the forecast trajectory. Additionally, during inference each member can be post-processed independently, which may reduce memory requirements when applying the method to very large real-time ensemble forecasts.

When applied to ECMWF IFS weekly mean $T_{2m}$ forecasts with the same objective of minimizing $\E[\mathrm{aCRPS}]$, the trajectory transformer successfully reduces systematic model biases whilst also improving or maintaining forecast reliability. Post-processed forecast anomalies exhibit spread-error ratios and total variance ratios close to one, which is not the case for the over-dispersive forecasts produced by the ensemble transformer. Crucially, these results are truly independent of both the ensemble size used to train the trajectory transformer (3 vs 9 members) and the ensemble size of the real-time forecast to be calibrated (9 vs 100 members). For this example application, there are limited impacts on the aCRPS of forecast anomalies compared to the benefits of bias correction. This likely reflects a combination of factors, including fundamental limitations of the model architecture, sub-optimal specification of input fields and/or training hyperparameters, non-stationarities in the reforecast training data, and the difficulty of calibrating already near-reliable IFS forecasts. However, we emphasize that the trajectory transformer is presented primarily as a proof-of-concept to demonstrate ensemble-size independence is achievable, rather than as a proposed solution for operational post-processing. Nevertheless, the trajectory transformer demonstrates several valuable properties: (i) true ensemble-size independence during both training and inference, (ii) maintained or improved reliability compared to raw forecasts, and (iii) successful bias correction without the systematic over-dispersion exhibited by ensemble transformers.

The results of this study have broader implications beyond the specific architectures considered, as the fundamental issue applies to any post-processing method that updates ensemble members based on the sample statistics of other members. In particular, we emphasize that reported sample CRPS estimates for post-processed ensembles should not be interpreted as definitive measures of forecast quality without independent verification of reliability.  The resulting deviations from reliability can be especially pronounced when minimizing the expected value of the (otherwise fair) aCRPS of a small ensemble (e.g. figure \ref{fig:expected_score_vs_var_ratio}).  The trajectory transformer described here represents one solution to this issue that eliminates member dependency entirely through architectural design. This approach sacrifices access to ensemble distribution information during inference for guaranteed ensemble-size independence. Other solutions may exist, though they are likely specific to the nature of the introduced dependency structure. For example, if the introduced dependency between members vanishes as $N \to \infty$ (as in the idealized member-by-member example of section \ref{section:idealized_example}), then it may be a justifiable compromise to use a sample CRPS estimator combined with large training ensembles if affordable/available. However, this argument does not extend to post-processing methods that introduce dependency between members that persists at large $N$ (i.e. $\Delta$ does not vanish as $N \to \infty$ in the notation of section \ref{section:intro}). In such cases, the underlying expected score does not target the proper integral CRPS (equation \ref{eq:crps_integral}) and the corresponding post-processed forecasts cannot be guaranteed to be reliable. Other alternative approaches could include adapted loss functions that enforce reliability constraints during training or derivation of generalized fair loss functions (if they exist) that explicitly account for an introduced dependency structure, which appears to be an open research question \citep{ferro2014fair}.

\section*{Acknowledgements}
The author thanks the PoET development team for making the source code available at \url{https://github.com/ecmwf-lab/poet/}, Jonathan Weyn for his help with understanding the PoET data loading interface, and Simon Lang, Jesper Dramsch, Zied Ben Bouall{\`e}gue, Matthew Chantry, Peter Deuben, Martin Leutbecher, and two anonymous reviewers for their thoughtful feedback on an earlier version of this work. During the preparation of this work, the author used generative AI tools (ChatGPT, OpenAI; Claude, Anthropic) to assist with drafting and debugging code, generating formatted \LaTeX{} tables from structured input data (e.g., YAML configuration files), and assistance with rephrasing and improving the clarity of selected passages in a preliminary manuscript draft. These tools were not used for study design, data analysis, or interpretation of results. All outputs were reviewed, validated, and edited by the author, who takes full responsibility for the content of the publication. Data from the ERA5 reanalysis are available to download from \url{https://www.ecmwf.int/en/forecasts/dataset/ecmwf-reanalysis-v5}. The IFS reforecasts used in this study are available from \url{https://apps.ecmwf.int/ifs-experiments/rd/hnll/}.

\section*{Conflict of interest}
The author declares no conflicts of interest.

\appendixpage
\begin{appendices}

\section{Linear calibration coefficients to minimize expected CRPS and aCRPS}
This appendix derives expressions for the linear calibration parameters $a$, $b$, and $c$, which minimize the expected values of CRPS and aCRPS when applied to the idealized Gaussian signal-plus-noise forecasts described in section \ref{section:idealized_example}.

\subsection{Ensemble forecast properties}
Before considering optimal calibration parameters, it is useful to define some forecast properties in terms of the underlying signal and noise characteristics, starting with the ensemble mean and deviations from the ensemble mean of raw forecasts:
\begin{equation}
    \bar{x}_j = s_j + \bar{n}_j, \quad \text{where } \bar{n}_j = \frac{1}{N}\sum_{k=1}^N n_{k,j} \sim \N\left(0, \frac{\alpha^2}{N}\right),
\end{equation}
\begin{equation}
    x_{k,j} - \bar{x}_j = n_{k,j} - \bar{n}_j.
\end{equation}
In the case of unbiased forecasts such that $\E[\bar{\hat{x}}_j - y_j ] = 0$ and thus $a=0$, the ensemble mean and deviation of calibrated forecasts can be written as follows: 
\begin{equation}
    \bar{\hat{x}}_j = b(s_j + \bar{n}_j),
\end{equation}
\begin{equation}
    \hat{x}_{k,j} - \bar{\hat{x}}_j = c(n_{k,j} - \bar{n}_j).
\end{equation}
It is then convenient to define the variance of the ensemble mean error for calibrated forecasts 
\begin{equation}
    \label{eq:error_var}
    \hat{\sigma}_\epsilon^2 \equiv \Var(y_j - \bar{\hat{x}}_j)  =  (1-b)^2\sigma_s^2 + \beta^2 + b^2\frac{\alpha^2}{N},
\end{equation}
where
\begin{equation}
    y_j - \bar{\hat{x}}_j = (1-b)s_j + e_j - b\bar{n}_j.
\end{equation}

\subsection{Objective functions and expected values}
The kernel representations of the CRPS and aCRPS are defined by equations \ref{eq:CRPS} and \ref{eq:aCRPS}. For the Gaussian signal-plus-noise data considered here, the key identity required to derive $\E[\mathrm{CRPS}]$ and $\E[\mathrm{aCRPS}]$ is the expected value of the folded normal distribution given by
\begin{equation}
\label{eq:folded_normal_identity}
\mathbb{E}|Z| = \sigma \sqrt{\frac{2}{\pi}} \exp\left(-\frac{\mu^2}{2\sigma^2}\right) + \mu \left(1 - 2\Phi\left(-\frac{\mu}{\sigma}\right)\right),
\end{equation}
where $Z \sim \N(\mu, \sigma^2)$ and $\Phi$ is the normal cumulative distribution function. When $\mu=0$ this reduces to the simpler form
\begin{equation}
\label{eq:folded_normal_identity_simple}
\mathbb{E}|Z| = \sigma \sqrt{\frac{2}{\pi}}.
\end{equation}

\subsubsection{Term 1: $\E|\hat{x}_{k,j} - y_j|$}
The difference between calibrated forecast members and observations can be written
\begin{equation}
    \hat{x}_{k,j} - y_j = (b-1)s_j - e_j + (b-c)\bar{n}_j + cn_{k,j}.
\end{equation}
The variance of this expression is defined below, which requires careful treatment of the terms $n_{k,j}$ and $\bar{n}_j$ as they are not independent. 
\begin{equation}
    \Var(\hat{x}_{k,j} - y_j) = (b-1)^2\sigma_s^2 + \beta^2 + \Var((b-c)\bar{n}_j + cn_{k,j}).
\end{equation}
This latter term can be expanded and simplified by noting that $\Cov(n_{k,j}, \bar{n}_j) = \frac{1}{N}\alpha^2$ to give
\begin{align}
    \Var(cn_{k,j} + (b-c)\bar{n}_j) &= c^2\alpha^2 + (b-c)^2\frac{\alpha^2}{N} + 2c(b-c)\frac{\alpha^2}{N}, \\
    &= \alpha^2\left[\frac{(N-1)c^2 + b^2}{N}\right].
\end{align}
This gives the following expression for the variance of forecast minus observation differences
\begin{align}
    \label{eq:var_xy}
    \hat{\sigma}_{XY}^2 \equiv \Var(\hat{x}_{k,j} - y_j) &= (1-b)^2\sigma_s^2 + \beta^2 + \alpha^2 \left[\frac{(N-1)c^2 + b^2}{N}\right],\\
                                                         &= \hat{\sigma}_{\epsilon}^2 + \left( \frac{N-1}{N} \right)\alpha^2c^2.
\end{align}
For an unbiased forecast, $\hat{\sigma}_{XY}^2$ can be inserted into equation \ref{eq:folded_normal_identity_simple} to give
\begin{equation}
    \label{eq:expected_dxy}
    \E|\hat{x}_{k,j} - y_j| = \sqrt{\frac{2}{\pi}\hat{\sigma}_{XY}^2} = \sqrt{\frac{2}{\pi}\left((1-b)^2\sigma_s^2 + \beta^2 + \alpha^2 \left[ \frac{(N-1)c^2 + b^2}{N}\right]\right)}.
\end{equation}

\subsubsection{Term 2: $\E|\hat{x}_{k,j} - \hat{x}_{l,j}|$}
The difference between two different calibrated ensemble members (i.e. $k \neq l$) is defined 
\begin{equation}
    \hat{x}_{k,j} - \hat{x}_{l,j} = c(n_{k,j} - n_{l,j}).
\end{equation}
Since $n_{k,j}$ and $n_{l,j}$ are independent, the variance of the difference between members is given by
\begin{equation}
   \label{eq:var_xx}
   \hat{\sigma}_{XX}^2 \equiv \Var(\hat{x}_{k,j} - \hat{x}_{l,j}) = \Var(c(n_{k,j} - n_{l,j})) = 2c^2\alpha^2.
\end{equation}
As ensemble members have the same expected value, $\hat{\sigma}_{XX}^2$ can also be inserted into equation \ref{eq:folded_normal_identity_simple} (i.e. $\mu=0$) to give
\begin{equation}
     \label{eq:expected_dxx}
    \E|\hat{x}_{k,j} - \hat{x}_{l,j}| = \sqrt{\frac{2}{\pi} \cdot 2c^2\alpha^2} = \frac{2|c|\alpha}{\sqrt{\pi}}.
\end{equation}

\subsubsection{$\E[\mathrm{CRPS}]$ of calibrated forecasts}
For sufficiently large $M$, taking the expectation over cases $j$ gives 
\begin{align}
    \E[\mathrm{CRPS}] &= \E \left[\frac{1}{N}\sum_{k=1}^{N}|x_k - y|\right] - \E\left[\frac{1}{2N^2}\sum_{k=1}^{N}\sum_{l=1}^{N}|x_k - x_l|\right],\\
                      &= \E|x_k - y| - \frac{N-1}{2N} \E|x_k - x_l|,\\
\end{align}
where there are $N(N-1)$ off-diagonal terms (i.e., $k \neq l$) in the double summation with $\E|x_k - x_l| \neq 0$. Substituting from equations \ref{eq:expected_dxy} and \ref{eq:expected_dxx} gives an expression in terms of the model and calibration parameters: 
\begin{equation}
    \label{eq:expected_crps}
    \boxed{\E[\mathrm{CRPS}] = \sqrt{\frac{2}{\pi}\hat{\sigma}_{XY}^2} - \frac{(N-1)|c|\alpha}{N\sqrt{\pi}},}
\end{equation}
where $\hat{\sigma}_{XY}^2$ is defined by equation \ref{eq:var_xy}.

\subsubsection{$\E[\mathrm{aCRPS}]$ of calibrated forecasts}
As above, taking the expectation over cases $j$ gives 
\begin{align}
    \E[\mathrm{aCRPS}] &= \E \left[\frac{1}{N}\sum_{k=1}^{N}|x_k - y|\right] - \E\left[\frac{1}{2N(N-1)}\sum_{k=1}^{N}\sum_{\substack{l=1 \\ l \neq k}}^{N}|x_k - x_l|\right],\\
                      &= \E|x_k - y| - \frac{1}{2} \E|x_k - x_l|.\\
\end{align}
Again, substituting from equations \ref{eq:expected_dxy} and \ref{eq:expected_dxx} gives an expression in terms of the model and calibration parameters: 
\begin{equation}
    \boxed{\E[\mathrm{aCRPS}] = \sqrt{\frac{2}{\pi}\hat{\sigma}_{XY}^2} - \frac{|c|\alpha}{\sqrt{\pi}}.}
\end{equation}

\subsection{Calibration parameters to minimize $\E[\mathrm{CRPS}]$}
The optimal calibration parameters are found by setting $\frac{\partial \E[\mathrm{CRPS}]}{\partial b} = 0$ and $\frac{\partial \E[\mathrm{CRPS}]}{\partial c} = 0$. Since $\hat{\sigma}_\epsilon^2$ (equation \ref{eq:error_var}) depends only on $b$ and not on $c$, the optimal value of $b$ can be determined independently of $c$, and the optimal value of $c$ can then be expressed as a function of the optimal $b$. To obtain the calibration parameters that minimize $\E[\mathrm{CRPS}]$, assume $c > 0$ and differentiate the objective function with respect to $c$, setting the derivative equal to zero. 
\begin{equation}
    \frac{\partial \E[\mathrm{CRPS}]}{\partial c} = \sqrt{\frac{2}{\pi}} \cdot \frac{1}{2\sqrt{\hat{\sigma}_{XY}^2}} \cdot \frac{\partial \hat{\sigma}_{XY}^2}{\partial c} - \frac{(N-1)\alpha}{N\sqrt{\pi}} = 0,
\end{equation}
where
\begin{equation}
    \frac{\partial \hat{\sigma}_{XY}^2}{\partial c} = \frac{2(N-1)c\alpha^2}{N}.
\end{equation}
Substituting and solving for $c$ gives
\begin{equation}
    \sqrt{\frac{2}{\pi}} \cdot \frac{(N-1)c\alpha^2}{N\sqrt{\hat{\sigma}_{XY}^2}} = \frac{(N-1)\alpha}{N\sqrt{\pi}},
\end{equation}
which simplifies to
\begin{equation}
    c = \frac{\sqrt{\hat{\sigma}_{XY}^2}}{\sqrt{2}\alpha}.
\end{equation}
Substituting equation \ref{eq:var_xy} and simplifying then yields the following expression for $c$ that minimizes $\E[\mathrm{CRPS}]$ for a specific value of $b$.
\begin{equation}
    \boxed{c^*_{\mathrm{CRPS}} = \sqrt{\frac{N}{N+1}} \cdot \frac{\hat{\sigma}_\epsilon(b)}{\alpha}}.
\end{equation}
Following the same procedure yields the optimal value for $b$, where 
\begin{equation}
    \frac{\partial \E[\mathrm{CRPS}]}{\partial b} = \sqrt{\frac{2}{\pi}} \cdot \frac{1}{2\sqrt{\hat{\sigma}_{XY}^2}} \cdot \frac{\partial \hat{\sigma}_{\epsilon}^2}{\partial b}.
\end{equation}
Since $\sqrt{\frac{2}{\pi}} \cdot \frac{1}{2\sqrt{\hat{\sigma}_{XY}^2}} > 0 $ for all $\hat{\sigma}_{XY}^2 > 0$, the optimal value of $b$ is found by setting $\frac{\partial \hat{\sigma}_{\epsilon}^2}{\partial b} = 0$
\begin{equation}
    \frac{\partial \hat{\sigma}_{\epsilon}^2}{\partial b} = \frac{2b\alpha^2}{N} -2(1-b)\sigma_s^2 =0,
\end{equation}
Since $\frac{\partial^2 \hat{\sigma}_{\epsilon}^2}{\partial b^2} = 2\left(\sigma_s^2 + \frac{\alpha^2}{N}\right) > 0$, this stationary point is a minimum. Solving gives
\begin{equation}
    \label{eq:optimal_b}
    \boxed{b^* = \frac{N\sigma_s^2}{N\sigma_s^2 + \alpha^2},}
\end{equation}
where $b^*$ is the ratio of signal variance to total variance of the ensemble mean. Substituting this expression for $b^*$ into equation~\ref{eq:error_var} gives the optimal value $\hat{\sigma}_\epsilon^*$, which is then used in the expression for $c^*$. 
\begin{equation}
    \hat{\sigma}_\epsilon^{*2} = \frac{\alpha^2\sigma_s^2}{N\sigma_s^2 + \alpha^2} + \beta^2.
\end{equation}

\subsection{Calibration parameters to minimize $\E[\mathrm{aCRPS}]$}
Since the derivatives of $\E[\mathrm{CRPS}]$ and $\E[\mathrm{aCRPS}]$ with respect to $b$ are the same, the optimal value of $b$ is the same for both CRPS and aCRPS objective functions. The optimal value of $c$ that minimizes $\E[\mathrm{aCRPS}]$ is again found by assuming $c > 0$ and differentiating with respect to $c$ and setting the derivative equal to zero. 
\begin{equation}
    \frac{\partial \E[\mathrm{aCRPS}]}{\partial c} = \sqrt{\frac{2}{\pi}} \cdot \frac{(N-1)c\alpha^2}{N\sqrt{\hat{\sigma}_{XY}^2}} - \frac{\alpha}{\sqrt{\pi}} = 0
\end{equation}
As above, substitution of equation \ref{eq:var_xy} and simplification yields an expression for $c$ that minimizes $\E[\mathrm{aCRPS}]$ for a specific value of $b$.
\begin{equation}
    \boxed{c^*_{\mathrm{aCRPS}} = \frac{N}{\sqrt{(N-1)(N-2)}} \cdot \frac{\hat{\sigma}_\epsilon(b)}{\alpha}}
\end{equation}

\section{The PoET ensemble transformer}
\label{appendix:ensemble_transformer}

In the PoET implementation of the ensemble transformer, the hidden state entering the transformer block at a given level is represented by
\begin{equation}
\label{eq:hidden_state_ensemble}
\mathbf{X} \in \mathbb{R}^{B \times N \times C \times H \times W},
\end{equation}
where $B$ is the batch size, $N$ is the ensemble size, $C$ is the channel depth, and $H,W$ are spatial dimensions. For $C_h$ attention heads, the linear value projection is represented by a $1 \times 1$ convolution such that

\begin{equation}
\label{eq:value_projection}
\mathbf{V} = \mathbf{W}^{(V)}*\mathbf{X} ,
\qquad
\mathbf{W}^{(V)} \in \mathbb{R}^{C \times C_h},
\end{equation}

where $\mathbf{W}^{(V)}$ represent the learnable weights and $*$ denotes a 2D convolutional operator over spatial locations with summation over input channels such that $\mathbf{V} \in \mathbb{R}^{B \times N \times C_h \times H \times W}$. The query and key projections are based on $3 \times 3$ strided convolutions represented by
\begin{equation}
\label{eq:query_key_projections}
\mathbf{Q} = \mathbf{W}^{(Q)}*\mathbf{X} ,
\qquad
\mathbf{K} = \mathbf{W}^{(K)}*\mathbf{X} ,
\end{equation}
where $\mathbf{W}^{(Q)} , \mathbf{W}^{(K)}  \in \mathbb{R}^{C \times C_h \times 3 \times 3}$ are learnable weights, $\mathbf{Q}, \mathbf{K} \in \mathbb{R}^{B \times N \times C_h \times H' \times W'}$, and $H', W'$ represent the reduced spatial dimensions after the strided convolution. For each head $c$, the dot-product attention weights between ensemble members $i$ and $j$ are calculated
\begin{equation}
\label{eq:attention_weights}
A_{c,i,j} = \mathrm{softmax}_i\left(  \frac{\sum_{x,y} Q_{i,c,x,y} K_{j,c,x,y}}{\sqrt{H'W'}} \right)
\end{equation}
where $\mathbf{A} \in \mathbb{R}^{C_h \times N \times N}$ and the softmax operation is applied over the member index $i$. The transformed value, $\hat{\mathbf{V}}$, for each member is then calculated as the sum of the original value and an attention-weighted combination of ensemble value perturbations
\begin{equation}
\label{eq:transformed_value}
\hat{V}_{j,c,x,y} = V_{j,c,x,y} + \sum_{i=1}^{N} A_{c,i,j} \left( V_{i,c,x,y}  - \overline{V}_{c,x,y} \right)
\end{equation}
where $\hat{\mathbf{V}} \in \mathbb{R}^{B \times N \times C_h \times H \times W}$ and $\mathbf{\overline{V}}$ represents the mean of $\mathbf{V}$ over the ensemble dimension. Finally, the updated values are projected back to the original channel dimension via a linear convolution and combined with a residual connection before ReLU activation
\begin{equation}
\label{eq:output_projection}
\mathbf{Z} = \mathrm{ReLU}\left(\mathbf{X} + \mathbf{W}^{(O)}*\hat{\mathbf{V}}\right),
\end{equation}
where $\mathbf{W}^{(O)} \in \mathbb{R}^{C_h \times C}$.

\end{appendices}
   
\newpage
\bibliographystyle{rss}
\bibliography{references}

\end{document}